\documentstyle[prl,multicol,aps]{revtex}
\begin{document}
\input{epsf.tex}
\epsfverbosetrue

\title{Four-wave mixing of linear waves and solitons
in  fibres with higher order dispersion}
\author{A.V. Yulin, D.V. Skryabin, and P.St.J. Russell}
\address{Department of Physics, University of Bath, Bath BA2 7AY, UK}
\date{\today}

\maketitle

\begin{abstract}
We derive phase-matching conditions for four-wave mixing
between solitons and linear waves in optical fibres with arbitrary dispersion
and demonstrate resonant  excitation of new spectral components
via this process.
\end{abstract}

\begin{multicols}{2}
\narrowtext

The use of any optical system, including fibres, for frequency
conversion  relies on the ability  to satisfy so-called
phase matching conditions, which critically depend on the
dispersive properties of the system, and on strong enough
nonlinear properties, to allow reduction in the threshold pump
power. The recent surge of interest in theoretical and
experimental studies of optical parametric processes in photonic crystal fibres (PCFs)
\cite{super,nature,fabio,harvey,opo1,herr,dima} is related to
their high nonlinearities, achieved by reduction in the core size
\cite{south}, and to the possibility of dispersion control by
suitable  design of  the photonic crystal cladding  \cite{nature}.

Here we study parametric generation of new frequencies resulting
from four-wave mixing (FWM) of solitons and continuous waves
(cw's) in optical fibres under  conditions where the effects of
 higher order dispersion are important, i.e., when pulses are
short and/or the frequency dependence of the group velocity
dispersion (GVD) is steep. Interest to this problem arises from
the fact that the Fourier components of a soliton
 are  dispersionless, while freely propagating
cw's are strongly dispersive. Therefore the phase-matching conditions
are expected to be satisfied at  frequencies different from those
 generated by the mixing of cw's. Addressing this issue is timely because of the availability of strongly
nonlinear small-core PCFs \cite{dima,south}, which  decrease the
threshold for observing of parametric processes by one to two
orders of magnitude compared to conventional fibres:
$\gamma_{conven}\sim 10^{-3}$W$^{-1}$m$^{-1}$, $\gamma_{pcf}\sim
10^{-2}$ to $10^{-1}$W$^{-1}$m$^{-1}$, where $\gamma$ is the nonlinear
fibre parameter \cite{agrawal1}. Strongly nonlinear PCFs have
already been used to demonstrate the coupling of solitons and cw
radiation  in: supercontinuum generation \cite{super,herr},
strong red and blue resonant, or Cherenkov, radiation from
solitons \cite{herr,dima} and cancellation of the soliton
self-frequency shift by the spectral recoil
\cite{dima}. Note, that it is natural to expect that FWM of
solitons and cw radiation is one of the many nonlinear processes
contributing to the shape  of the supercontinuum spectra.

The problem of mixing of  solitons and cw using the idealized
nonlinear Schr\"odinger equation (NLS), i.e., with higher order
dispersion disregarded, has been analyzed in a number of papers in
the past \cite{hasegawa,haus,gordon,kaminis,akhmed,park}. In these
cases several exact analytical solutions for a soliton sitting on
the cw background were found \cite{akhmed,park} and different
perturbation techniques  suggested
\cite{hasegawa,haus,gordon,kaminis}. However, none of these studies
have  addressed the issue of generation of new frequencies by
FWM of  solitons and cw light -- the central focus of
this paper.

We  assume  that the dynamics of the dimensionless amplitude
$A(t,z)$ of the fundamental fiber mode is governed by the
generalized NLS equation \cite{nature}
\begin{eqnarray}
&&\partial_{z}A=iD(i\partial_t)A+
iA\int_{-\infty}^{+\infty}{R(t')|A(t-t',z)|^{2}}dt'.\label{eqn1}
\end{eqnarray}
The dispersion operator in Eq. (\ref{eqn1}) is given by
\begin{equation}
D(i\partial_{t})\equiv \sum_{m= 2}^M
{\tau^{2-m}\partial_{\omega}^{m-2}\beta_2(\omega_0) \over
m!|\beta_2(\omega_0)|} [i\partial_{t}]^m, \label{eq_disp}
 \end{equation}
where $\tau$ is the pulse duration and $\omega_0$ is the reference
frequency. To avoid any ambiguity in the analytical expressions we
adopt the convention of using round brackets $(..)$  to indicate
the arguments of  functions or operators and $[..],~\{.. \}$ for
all other purposes. $R(t)$ is the response function of the
material, which includes instantaneous Kerr and delayed Raman
nonlinearities:
\begin{equation}
R(t)=[1-\theta]\Delta(t)+\theta \alpha
\Theta(t)e^{-t/\tau_2}\sin(t/\tau_1) \label{eqn2}\end{equation}
Here $\Delta(t)$ and $\Theta(t)$ are, respectively, delta and
Heaviside functions,
$\alpha= [\tau_1/\tau_2+\tau_2/\tau_1]/\tau_2$, $\theta=0.18$,
$\tau_1=12.2fs/\tau$, $\tau_2=32fs/\tau$ \cite{agrawal1}. $t$ is
the time   in the reference frame moving with the group velocity
$v_0=v(\omega_0)$ and measured in the units of $\tau$:
$t=[T-z/v_0]/\tau$, where $T$ is the physical time. $z=Z/L_{gvd}$,
where $Z$ is  the distance along the fiber and
$L_{gvd}=\tau^2/|\beta_2(\omega_0)|$ is the  GVD length. The field
amplitude  $A$ is  measured in  units of $N/\sqrt{\gamma
L_{gvd}}$, where $N^2$ is the ratio of the peak power of the pump
pulse to the peak power  of a fundamental soliton with duration
$\tau$.

The dispersive properties of the  soliton and linear cw are
crucial for the following, so we now discuss them  in some detail.
Looking for a linear wave solution of (\ref{eqn1}) in the form
$A\sim e^{iD_xz-i\delta_xt}$ we find $D_x=D(\delta_x)$. In what
follows the subscript $x$ can take any convenient notation.
E.g., $\delta_s$ and $\delta_{cw}$ correspond below to the frequency shifts
of the soliton and cw-pump, respectively.
Thus the physical wave number of a linear wave with frequency
$\omega_0+\delta_x/\tau$ is given by
$k_x=k(\omega_0)+D_x/L_{gvd}$. Plotting $k_x-k(\omega_0)$ vs
$\delta_x$ one  simply recovers the dispersion profile of the
fibre. The single soliton solution
\begin{equation}
 A=F(\xi)e^{-i\delta_st+i[D_s+q]z},~
F(\xi)=\sqrt{2q}sech(\xi/r) \label{eqn3}
\end{equation}
satisfies Eq. (\ref{eqn1}) if all derivatives of the function
$F(\xi)$ higher than second are disregarded, $\theta=0$, and
$D_s^{\prime\prime}<0$, i.e., the GVD at the soliton frequency is
anomalous. Here $\xi=t-D^{\prime}_sz$,
$r=\sqrt{-D^{\prime\prime}_s/( 2q)}$, $^{\prime}$ denotes the
derivative with respect to $\delta$ and $q>0$ is the additional
shift of the soliton wave number. Representing $F(\xi)$ through
the inverse  transform of its Fourier image $\tilde F(\delta)$,
i.e. $F=\int d\delta \tilde Fe^{i\xi[\delta_s-\delta]}$, one
finds that the wave number of a Fourier component of the soliton
with frequency $\omega_0+\delta_x/\tau$ is given by
$k_{s/x}=k(\omega_0)+\{D_s+q-[\delta_s-\delta_x]D^{\prime}_s\}/L_{gvd}$.
Thus plotting $k_{s/x}-k(\omega_0)$ vs $\delta_x$ will give the tangent line
to the curve $k_x-k(\omega_0)$ taken at the point
$\delta_x=\delta_s$ and shifted up by $q$. The linear dependence
of $k_{s/x}$ on $\delta_x$ is a reflection of the
dispersion-free nature of solitons.

We  seek solutions of Eq. (\ref{eqn1}) in the form
\begin{equation}
A=\{F(\xi)+g(z,\xi)\}e^{iz[D_s+q]-i\delta_s t}, \xi\equiv t-z
D^{\prime}_s.\label{anzats}
\end{equation}
Assuming that $g$ is  a  quasi-linear wave we derive
\begin{equation}
i p=\partial_z g-i\tilde D(i\partial_{\xi})
g-i2F^2g-iF^2g^*,\label{eqn4}
\end{equation}
where
$\tilde D(i\partial_{\xi})=-q-D_s-iD_s^{\prime}\partial_{\xi}-D(i\partial_{\xi}+\delta_s)$
and $p=[D(i\partial_{\xi}+\delta_s)-D_s-iD_s^{\prime}\partial_{\xi}
+1/2D_s^{\prime\prime}\partial^2_{\xi}]F$.
Now we split $g$ into two parts
\begin{equation}
g=we^{i\phi}+\psi,~
\phi=z\tilde D(\delta_{cw}-\delta_s)+\xi[\delta_s-\delta_{cw}]
\label{gpsi}
\end{equation}
The $w$-term in  Eq. (\ref{gpsi}) is the
weak cw pump, which obeys Eq. (\ref{eqn4}) with $p=F=0$.  The
$\psi$-term is the field generated through the mixing of the soliton and cw pump
 and by the soliton itself.
Seeking  $\psi$ in the form $\psi=\psi_p+\psi_+e^{iz\tilde
D(\delta_{cw}-\delta_s)}+\psi_-^* e^{-iz\tilde
D(\delta_{cw}-\delta_s)}$ we find that $\psi_{p}$ obeys
\begin{equation}
i \left[\begin{array}{cc}p\\-p^*\end{array}\right]=
\left\{\partial_z +i\hat{\cal L}\right\}
\left[\begin{array}{cc}\psi_p\\\psi^*_p\end{array}\right],\label{eq_a}
\end{equation}
and $\psi_{\pm}$ are governed by the system of coupled equations
\begin{eqnarray}
iF^2we^{i\xi[\delta_s-\delta_{cw}]}
\left[\begin{array}{cc}2\\-1\end{array}\right]=
\left\{\partial_z +i\hat{\cal L}+i\hat{\cal W}\right\}
\left[\begin{array}{cc}\psi_+\\ \psi_-\end{array}\right],\label{eq_c}
\end{eqnarray}
where
\begin{eqnarray}
\nonumber && \hat{\cal L}=\left[\begin{array}{cccc}-\tilde D(i\partial_{\xi})-2F^2&-F^2
\\F^2&\tilde D(-i\partial_{\xi})+2F^2\end{array}\right],\\
\nonumber  && \hat{\cal W}=\tilde D(\delta_{cw}-\delta_s)\left[\begin{array}{cccc}1&0
\\0&1\end{array}\right].
\end{eqnarray}
Eq. (\ref{eq_a}) does not depend on the cw-pump and it is  known
to describe the emission of non-localized dispersive waves from
solitons --  so-called resonant or Cherenkov radiation -- in
fibres with zero GVD points \cite{herr,dima,wai1,karpman1,karl}.
 This radiation exists
because the continuous part of the spectrum of the operator
$\hat{\cal L}$  has a zero eigenvalue,
which ensures   resonance with the forcing term $p$.
The operator $\hat{\cal L}+\hat{\cal W}$ is expected to have
several different resonances, a situation which to our
knowledge has not previously been considered. These resonances are
driven by FWM between the soliton and the cw, see the left-hand
side in Eq. (\ref{eq_c}). To find continuous spectra of $\hat{\cal
L}$ and $\hat{\cal L}+\hat{\cal W}$ we neglect the $F^2$-terms and
look for  eigenfunctions in the form $\psi_p\sim e^{i\kappa
z+i[\delta_s-\delta]\xi}$ and $\psi_{\pm}\sim e^{\pm i\kappa z\pm
i[\delta_s-\delta]\xi}$ Assuming that the wave numbers of the
continuum modes are matched with the wave numbers of the driving
terms in Eqs. (\ref{eq_a},\ref{eq_c}), i.e. $\kappa=0$, we find:
\begin{eqnarray}
q+D_s-[\delta_s-\delta]D_s^{\prime}=D(\delta),&&\label{eq_a1}\\
\nonumber \pm D_{cw}\mp \{q+D_s-[\delta_s-\delta_{cw}]D_s^{\prime}\}+&&
\\\{q+D_s-[\delta_s-\delta]D_s^{\prime}\}
=D(\delta)&&\label{eq_a2}.
\end{eqnarray}
Eqs. (\ref{eq_a1},\ref{eq_a2}) are the equations for $\delta$. The
roots of these equations, which can be easily found graphically,
yield the frequencies of the emitted radiation.  The right-hand
side $D(\delta)$ is simply the dispersion of linear waves, which
is a nonlinear function of $\delta$. The left-hand sides are
straight lines. Fitting the experimentally measured dispersion
characteristics of the fibres usually results in a very high order
polynomial for $D(\delta)$
 \cite{nature,dima}. Here, however we
use the simplest illustrative example, when only the third order
dispersion is included, i.e.
$D(\delta)=-\delta^2/2+\epsilon\delta^3$,
$\epsilon=\partial_{\omega}\beta_2(\omega_0)/(6\tau|\beta_2(\omega_0)|)$.
This model accurately describes  regions where the frequency dependence
of $\beta_2(\omega)$ is quasi-linear - typical for
telecom and PCF fibres, see, e.g., \cite{dima}. Figs. 1(a) and (b)
show  plots of the right- and left-hand sides of Eqs.
(\ref{eq_a1},\ref{eq_a2}). Eq. (\ref{eq_a1}) gives the Cherenkov
resonance \cite{herr,dima,wai1,karpman1,karl}, which does not
depend on the cw-pump. The roots resulting from  Eqs.
(\ref{eq_a2}), however, are determined by the cw-pump. Depending
on the value of $\delta_{cw}$ Eqs. (\ref{eq_a2}) produce either
two or four new roots, see Figs. 1(a),(b). Assuming that
$\delta=\delta_{cw}$, we find that  Eq. (\ref{eq_a2}) with $+$ is
satisfied, which means that one of the new roots always coincides
with $\delta_{cw}$. Clearly the position of these new resonances
strongly depends on, and can be controlled by, the frequency of
the cw-pump and by engineering the linear dispersion profile
 of the fibres.

The four-wave mixing nature of the new resonances becomes clear if
we make use of the expressions for the wave numbers of the
dispersive waves and the Fourier components of the soliton (see
the discussion around Eq. (\ref{eqn3})). Indeed, Eqs.
(\ref{eq_a2}) can be rewritten in the form
\begin{eqnarray}
 \pm[k_{cw}-k_{s/cw}]+k_{s/rad}=k_{rad}.\label{eq_b2}
\end{eqnarray}
Here $k_{rad}$, $k_{cw}$ are wavenumbers of the resonant radiation
and the cw-pump, and $k_{s/rad}$, $k_{s/cw}$ are the
wavenumbers of the Fourier components of the soliton at the
frequencies of the resonant radiation and the cw-pump.

To confirm our analytical findings we have carried out a series of
numerical experiments with parameters close to the ones in
\cite{dima}. Comparing Figs. 1 (a),(b) with  1 (c),(d) one  sees
excellent agreement between the analytical predictions and the
frequencies emerging from the modelling  of (\ref{eqn1}). The
efficiency of excitation of the new frequencies strongly depends
on the choice of $\delta_s$ and $\delta_{cw}$ relative to each
other and to the zero GVD point. This explains why not all the
resonances are observed simultaneously. Changing $\delta_{cw,s}$
we have been able to observe all of the newly predicted
resonances. Theoretical analysis of this problem is possible
within the framework of  Eq. (\ref{gpsi}) and we leave it for
a future study. The cw powers required to observe the new FWM
resonances are of the order or less than $1$W, see the Fig. 1
caption.

\begin{figure}
\setlength{\epsfxsize}{8.8cm} \centerline{\epsfbox{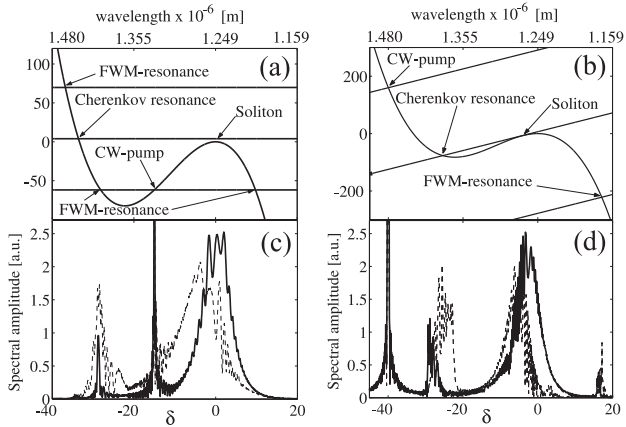}}
\caption{
 Resonance frequencies (a,b) calculated from Eqs. (\ref{eq_a1},\ref{eq_a2})
and  supporting numerical modelling (c,d). Straight/curved lines
in (a,b) are the left/right-hand sides of Eqs.
(\ref{eq_a1},\ref{eq_a2}). Choosing $\omega_0=2\pi\times 240$THz
and the fibre parameters from [7] we find $\beta_2=-47$ps$^2/$km
and $\partial_{\omega}\beta_2=-0.7$ps$^3/$km. $\delta_s=0$,
$w=0.1$, $\delta_{cw}=-15$ for (a,c) and $\delta_s=-4$, $w=0.3$,
$\delta_{cw}=-40$ for (b,d). Other parameters are $\tau=170$fs,
$q=4$, $N=1.5$. For $\gamma=0.05$(Wm)$^{-1}$ this corresponds to
the pulse peak power $225$W, the cw-power $0.12$W for (c) and
$1.2$W for (d). The full lines in (c,d) corresponds to $\theta=0$,
i.e. Raman is off, and the dashed  lines to $\theta=0.18$, i.e.
Raman is on. The propagation distance is $2.8$m (c) and $16$m (d).
The top axes are marked in wavelength units.}
\end{figure}

We have analyzed FWM between solitons and cw-pump in fibres with
higher order dispersion and predicted the generation of new
frequencies, which can be controlled by tuning the cw-pump. In
addition to their fundamental significance,  our results  have
important implications for the generation of laser light and
broad-band supercontinua  in  new frequency ranges.

This work is partially supported by the EPSRC grant GR/S20178/01.

\end{multicols}
\end{document}